
%
\magnification=\magstep1
\vsize=23truecm
\hsize=15.5truecm
\hoffset=.2truecm
\voffset=.8truecm
\parskip=.2truecm

\font\ti=cmbx10 scaled\magstep1

\def\br{\hfill\break\noindent}

\def \ot {\otimes}
\def \g5{\gamma_5}

\def \l4{\Bigl( {\rm Tr}( KK^*)^2-({\rm Tr}KK^*)^2\Bigr)}
\def \bp{\oplus }
\def \k2{{\rm Tr}KK^*}
\def \slash#1{/\kern -6pt#1}
\def \di{\slash{\partial}}

\def \bt{\otimes}
\pageno=0
%
%
\baselineskip=.5truecm
\footline={\hfill}
{\hfill ETH/TH/95-28}
\vskip.2truecm
{\hfill 23 October 1995}
\vskip2.1truecm
\centerline{\ti The Scalar Potential in Noncommutative Geometry }
\vskip1.2truecm
\centerline{  A. H. Chamseddine }
\vskip.8truecm
\centerline{ Theoretische Physik}
\centerline{ETH-H\"onggerberg}
\centerline{CH-8093, Z\"urich}
\centerline{Switzerland}
\vskip1.2truecm

\centerline{\bf Abstract}
\vskip.5truecm
\noindent
We present a derivation of the general form of  the scalar potential
in  Yang-Mills theory of a non-commutative space which is a product
of a four-dimensional manifold times a discrete set of points. We show
that a non-trivial potential without flat directions is obtained after
eliminating the auxiliary
fields only if constraints are imposed on the mass matrices
utilised in the Dirac operator. The constraints and potential are
related to a prepotential function.
\vfill

\eject
\noindent
One of the most attractive aspects of noncommutative geometry [1-2]
is that it deals with spaces which could not be handled otherwise.
 It also offers tools which could be used in probing the hidden structure
of space-time [3]. An encouraging indication of the relevance of
this new geometry to particle physics is the fact that the simplest
noncommutative space one considers, where the algebra is
$C^{\infty }(M)\ot (M_1(C)\bp M_2(C)\bp M_3(C) )$ reproduces, classically,
the standard model with all of its details [4-5]. It also provides an
appealing geometrical picture for the Higgs field. These ideas were
then applied to grand unified theories [6] and it was found that the
main advantage of this scheme is that the composition of the
Higgs sector is almost
uniquely fixed. This is in complete contrast to the usual treatment
of grand unified theories where the advantages  gained in
reduction of the fermionic representations and the unification
of the coupling constants are wiped out by the arbitrariness in the
Higgs fields representations. What is still missing in noncommutative
geometry is the identification of
the hidden new symmetries present as this is  needed
to protect, at the quantum level, any classical predictions [7].
A step in this
direction has  recently been made where it was conjectured that the
hidden symmetry of the noncommutative standard model is the
quantum group $SU(2)_q$ at the cubic root of unity [8].

It is therefore important to simplify the calculations involved
 in deriving  a general formula for the Yang-Mills noncommutative
action associated with a product space
of a continuous manifold times a discrete set of points, with
an appropriate fermionic representation. This program was
started in [6] but the result involved auxiliary fields
which have to be  eliminated. This last step was only carried
out for some specific models, and it was found that the
form of the potential depends on the mass matrices used in
the Dirac operator. These mass matrices also turn out to be the vevs (vacum
expectation values ) of the Higgs fields and must satisfy
certain conditions. The purpose of this letter is to
derive a general form of the potential, after elimination
of the auxiliary fields, and to determine the constraints that must be
imposed on the mass matrices in the Dirac operator to obtain a physically
acceptable potential.

We start with the spectral
triple $({\cal A},h,D)$,
where $h$ is a Hilbert space, ${\cal A}$ is an involutive algebra
of operators on $h$, and $D$ is an unbounded self-adjoint
operator on $h$ [1].
Let $M$ be a compact Riemannian spin-manifold,
${\cal A}_1$ the algebra of functions on $M$, and
$(h_1,D_1,\Gamma_1 )$ the Dirac-$K$ cycle  with $h_1\equiv L^2 (M,
\sqrt g d^d x )$ on ${\cal A}_1$. Let
$({\cal A}_2,h_2,D_2)$ be given by ${\cal A}_2=\bp_{p=n_1}^{n_N}
M_{p}(C) $, where $M_p(C)$ is the set of all
$p\times p $ matrices and $h_2=\oplus_{p=n_1}^{n_N} h_{2,p} $
where $h_{2,p}$  is the Hilbert
space $C^p$.
We take ${\cal A}$ and $D$  to be
$$\eqalign{
{\cal A}&={\cal A}_1\otimes {\cal A}_2 \cr
D&=D_1\otimes 1 +\Gamma_1 \otimes D_2 .\cr}\eqno(1)
$$
To every $ a\in {\cal A}$ we associate a N-plet $(a_{n_1} \cdots
a_{n_N})$ of matrix-valued functions on $M$, where $a_p$ are
 $p\times p$ matrix valued functions.
In this decomposition, the operator $D$ becomes
$$
 D=\pmatrix {\di\bt 1_{n_1}\bt 1_3&\g5 \bt M_{12}\bt K_{12}
&\ldots &\g5 \bt M_{1N}\bt K_{1N} \cr
\g5 \bt M_{21}\bt K_{21} &\di \bt 1_{n_2}\bt 1_3 &\ldots &\g5 \bt
M_{2N}\bt K_{2N}\cr
\vdots &\vdots &\ddots &\vdots \cr
\g5 \bt M_{N1}\bt K_{N1} &\g5 \bt M_{N2}\bt K_{N2} &\ldots & \di
 \bt 1_{n_N} \bt 1_3 \cr},\eqno(2)
$$
where $M_{mn}^* =M_{nm}$ and $m,n=1,\cdots ,N, m\not=n$  and the
$K_{mn}$ are generation mixing $3\times 3$ matrices.

Let $E$ be a vector bundle characterized by the vector space
$\cal E$ of its sections.  We shall consider the example where
${\cal E}={\cal A}$.
 Let $\rho $ be a self-adjoint element
in the space, $\Omega^1 ({\cal A})$, of one forms
$$
\rho =\sum_{i} a^i db^i  ,\eqno(3)
$$
where $\Omega^. ({\cal A})=\oplus_{n=0}^{\infty} \Omega^n ({\cal
A})$ is the universal differential algebra, with $\Omega^0
({\cal A})={\cal A}$ [1].
An involutive representation of $\Omega^. ({\cal A}) $ is provided by the map
$\pi : \Omega^. ({\cal A})\rightarrow B(h) $  defined by
$$
\pi (a_0da_1...da_n)=a_0[D,a_1][D,a_2]...[D,a_n],  \eqno(4)
$$
where $B(h)$ is the algebra of bounded operators
on $h$.
The image of the one-form $\rho $ is
$$
\pi (\rho ) =\sum_i a^i[D, b^i],  \eqno(5)
$$
and this takes the matrix form:
$$
\pi (\rho )=\pmatrix{\gamma^{\mu}\bt A_{\mu 1} \bt 1_3 &\g5 \bt
\phi_{12} \bt K_{12}
&\ldots &\g5 \bt \phi_{1N} \bt K_{1N}\cr
\g5 \bt \phi_{21} \bt K_{21}& \gamma^{\mu}\bt A_{\mu 2} \bt 1_3
&\ldots &\g5 \bt \phi_{2N}
\bt K_{2N} \cr
\vdots &\vdots &\ddots &\vdots \cr
\g5 \bt \phi_{N1} \bt K_{N1}& \g5 \bt \phi_{N2}\bt K_{N2}
 &\ldots &\gamma^{\mu}\bt A_{\mu N} \bt 1_3 \cr} ,\eqno(6)
$$
where the new variables $A_{\mu m}$ and $\phi_{mn} $ are functions of the
$a^i$ and  $b^i$ given by
$$\eqalign{
A_{\mu m} &=\sum_i a_m^i\partial_{\mu} b_m^i, \qquad m=1,2,\ldots , N,\cr
\phi_{mn}&=\sum_i a_m^i (M_{mn}b_n^i -b_m^iM_{mn}),\qquad
m\not= n, \cr}\eqno(7)
$$
and satisfy $A_{\mu m}^{\ast }=-A_{\mu m}$ and $\phi_{mn}^{\ast }=\phi_{nm} $.
The two-form $d\rho $ is:
$$
d\rho =\sum_i da^i db^i  \eqno(8)
$$
and its image under the involutive representation $\pi $ is
given by
$$
\pi (d\rho )=\sum_i [D, a^i][D, b^i] . \eqno(9)
$$
The curvature $\theta $  is defined by
$$
\theta =d\rho +\rho^2  .\eqno(10)
$$
It must be noted  that the representation $\pi $ is
ambiguous [4], a fact that will explain the appearence of auxiliary
fields. This can be seen from the fact that if $\pi (\rho )$ is set
to zero, $\pi (d\rho ) $ is not necessarily zero, and the correct
space of forms to work on is $\Omega_D^. ({\cal A})=\oplus \Omega_D^n
({\cal A})$, where
$\Omega_D^n ({\cal A}) ={\Omega^n ({\cal A}) \over ({\rm Ker} \pi +
d {\rm Ker} \pi )_n }$, where ${\rm Ker} \pi $ is the kernel of the map $\pi $.

The representation of the curvature
$\pi (\theta )$ can be written in terms of components. First, the
diagonal elements are given
$$
\pi (\theta )_{mm}={1\over 2}\gamma^{\mu \nu}F_{\mu\nu}^m
+\Bigl(\sum_{p\not=m} (\vert K_{mp}\vert^2\bigl( \vert H_{mp}
\vert ^2 -\vert M_{mp}\vert^2 \bigr)
 \Bigr)-X'_{mm} \qquad m=1,2\ldots ,N, \eqno(11)
$$
where we have defined
$$\eqalign{
X'_{mm}&=\sum_i a_m^i\di^2 b_m^i +\ (\partial^{\mu}
A_{\mu}^m+A^{\mu m}A_{\mu}^m)
 +\bigl[ \sum_{n\not =m}a_m^i \vert K_{mn}\vert^2
M_{mn}M_{nm}, \ b_m^i\bigr] \bigr) \cr
F_{\mu\nu}&=\partial_{\mu}A_{\nu}^m-\partial_{\nu}A_{\mu}^m
+[A_{\mu}^m,A_{\nu}^m]  \cr
H_{mp}&= \phi_{mp} +M_{mp}. \cr}\eqno(12)
$$
The non-diagonal elements of $\pi (\theta )$ are given by
$(m\not=n )$:
$$\eqalign{
\pi (\theta )_{mn}&=-\g5 K_{mn}\bigl( \di H_{mn}+ A_m H_{mn}
-H_{mn}A_n\bigr) \cr
&\qquad +\sum_{p\not= m,n}  K_{mp}K_{pn}
\bigl( H_{mp}H_{pn} -M_{mp}M_{pn}\bigr)-X_{mn} .\cr}\eqno(13)
$$
The curvature $\theta $ is self-adjoint:
$\pi (\theta )_{mn}^*=\pi (\theta )_{nm}$.

The full noncommutative action is given by
$$\eqalign{
I&=\int d^4x \Bigl(- {1\over 4}\sum_m F_{\mu \nu}^m F^{\mu\nu m}\cr
&\qquad +\sum_{p\not= m}\vert \partial_{\mu}H_{mp} +A_{\mu m}
H_{mp}-H_{mp}A_{\mu p}\vert^2  -V \Bigr) \cr
& +\Bigl( \Psi , (D+\pi (\rho ))\Psi \Bigr) ,\cr}
\eqno(14)
$$
where the fermions are collectively denoted by $\Psi $, and the
scalar potential by $V$.
The fields  $X_{mn}'$ and $X_{mn}$ are not all independent,
and the relations among them   depend on the structure of
the mass matrices $M_{mn}$.
We would like to find out what sort of constraints  must be
imposed on  the mass matrices in order to get a non-trivial
scalar potential. In the generic case where these matrices are arbitrary,
and the dimensions of all the matrices are different than one,
every term contributing to the potential in the curvature will be moded out.
To see this explictely, we note that the potential is given by
$$\eqalign{
V&=
\Bigl\vert  \sum_{p\not= m}\vert K_{mp}\vert^2
\bigl( \vert H_{mp}\vert^2 -\vert M_{mp}\vert^2 \bigr)
-X'_{mm})\Bigr\vert^2 \cr
& + \sum_{n\not=m}\sum_{p\not=m,n}\Bigl\vert
K_{mp}K_{pn}\bigl(
H_{mp}H_{pn}-M_{mp}M_{pn}\bigr)-X_{mn}\Bigr\vert^2
\Bigr),\cr}\eqno(15)
$$
which is a sum of squares. Each of these terms contain an auxiliary
field, and when all are eliminated   the potential vanishes. Therefore,
for the matrices $M_{mn}$ to correspond to minima of the potential
they must sastisfy certain conditions. We shall now determine the
necessary constraints.

The space of auxiliary fields is found by calculating the kernel
of the operator $\pi (d\rho )\vert_{\pi (\rho )=0}$. A simple calculation
gives
$$\eqalign{
({\rm Aux})_{mm}&=\sum_i a_m^i\di^2 b_m^i
 +\bigl[ \sum_{n\not =m}a_m^i\vert K_{mn}\vert^2
M_{mn}M_{nm}, \ b_m^i\bigr]  \cr
({\rm Aux})_{mn}&= \sum_i a_m^i \sum_{p\not= m,n}
K_{mp}K_{pn}\bigl( M_{mp}M_{pn}
b_n^i -b_m^iM_{mp}M_{pn} \bigr) ,\qquad m\not= n,\cr}\eqno(16)
$$
Equation (16) is a function of the square of the matrices
$M_{mn}$ and as these in general are not equal to a linear
combination of themselves, the auxiliary fields produced
will be nonconstrained scalar functions. When the curvature
two-form is moded out by the auxiliary fields along diagonal
and non-diagonal terms, the scalar parts of the curvature drop
out. Therefore there is a need to carefully chose the matrices
$M_{mn}$ in order to get an acceptable scalar potential.

The simplest possibility corresponds to the case where the dimension
of one of the matrices, e.g. $n_1$ is equal to one.  Then the
commutator in $({\rm Aux})_{11}$ drops out, and it reduces to
$({\rm Aux})_{11}=\sum_i a_1^i\di^2 b_1^i $. After elimination of this field,
the contribution of the first diagonal part of the
curvature to the potential reduces to the form:
$$
v_{11}=\sum_{p\not= 1}\bigl( \vert K_{1p}\vert^2 \bigr)^{\perp}
\bigl( \vert H_{1p} \vert^2 -\vert M_{1p}\vert^2
\bigr) ,
$$
where
$$\bigl( \vert K_{mp}\vert^2\bigr)^{\perp} =\vert K_{mp}\vert^2
-{\rm Tr}\bigl(\vert K_{mp}\vert^2\bigr) 1_3 , \eqno(17)
$$
and where we have normalised the trace so that Tr$(1)=1$.
Therefore, whenever the dimension of one of the matrix algebras is
one, there is no need to impose any conditions
on the mass matrices $M_{1p}$ for the potential to survive. This is the
situation encountered in the standard model where the dimension of one of the
matrix algebras is one.

In all other cases we have to restrict the choice of the mass matrices
so that some of the auxiliary fields could be expressed in terms
of the Higgs fields, and this would make them linearly dependent.
As the potential is the sum of squares, its minimum
occurs when each term vanishes. Each of these terms
is quadratic in the fields, so  the minimum values of the Higgs fields
$H_{mp}$ must satisfy  quadratic equations.

First, we have to assume that the generation mixing matrices
are related. The simplest possibility  is
$$
\vert K_{mn}\vert^2 =\vert K\vert^2 \qquad \forall m, n ,\eqno(18)
$$
This condition could be slightly relaxed, but the analysis will
become less transparent.
The most general constraint on the matrices $M_{mn}$ is then:
$$
\sum_{p\not= m \not= n} M_{mp}M_{pn} =\alpha_{mn} M_{mn}. \eqno(19)
$$
Using  equations (18) and (19)
it is  easy to show that the fields $X_{mn}$ reduce to
$$
X_{mn}=-\vert K\vert^2 \alpha_{mn}\phi_{mn}, \eqno(20)
$$
 and the
terms $v_{mn}$ in the potential (the potential is a function of the
square of  $v_{mn}$) simplify to  the nice form
$$
v_{mn}=\vert K\vert^2 \bigl( \sum_{p\not= m\not= n} H_{mp}H_{pn}
-\alpha_{mn}H_{mn} \bigr) . \eqno(21)
$$
It is clear that the equations of motion satisfied by $H_{mn}$
are the same as that satisfied by the mass matrices $M_{mn}$.
In other words the
fields $\phi_{mn} $ have zero vevs. The other contributions to the potential
 from $v_{mm}$ do not survive as there is no possible simplification  for
$X_{mm}'$. When $X_{mm}'$ is eliminated the  parts coming
from the diagonal components of the curvature  disappear.
Therefore, except when ${\rm dim}[n_m] =1$, the terms in the
potential dependent on $v_{mm}$ vanish.
We can understand this result by studying
the gauge transformations of the Higgs fields.
{}From the gauge transformations
of the curvature
$$
\theta \rightarrow {^g}\theta =g\theta g^{\ast}, \eqno(22)
$$
where $g$ is an element of $\cal A $ with the representation
$$
g\rightarrow {\rm diag} (g_1, \ldots , g_N), \eqno(23)
$$
we can show that
$$
^g H_{mn}=g_m H_{mn}g_n^{\ast} \qquad m\not= n ,\eqno(24)
$$
Equation (24) then implies that
$$
\sum_p H_{mp}H_{pn} \rightarrow g_m \bigl(\sum_p H_{mp}H_{pn}
\bigr)g_n^{\ast} ,\eqno(25)
$$
and transforms in the same way as $H_{mn}$  allowing for
constraints to be imposed relating $\sum_p H_{mp}H_{pn}$ to $H_{mn}$.
 On the other hand no such relation
is possible for $\sum_p H_{mp}H_{pm}$ as this combination transforms
as
$$
g_m \bigl( \sum H_{mp}H_{pm}\bigr) g_m^{\ast} , \eqno(26)
$$
and there is no Higgs field with such a transformation.
We already noted that the potential is a sum of squares, a feature
which is also present in globaly supersymmetric theories. There, one can
define a gauge invariant superpotential $g(z^i)$, holomorphic
in the complex scalar fields $z^i$, and the potential takes the very
elegant form
$$
V=\Bigl\vert {\partial g(z) \over \partial z^i}\Bigr\vert^2 .\eqno(27)
$$
By analogy, we can define here a prepotential gauge invariant function
$$
g(H)={\rm tr} \bigl( {1\over 3} \sum_{p\not= m \not =n }
H_{mp}H_{pn}H_{nm} -{1\over 2} \sum_{m,n} \alpha_{mn} H_{mn}H_{nm}
\bigr) \bigr) .\eqno(28)
$$
The total potential in this case can be
written in the simple form:
$$
V =\bigl({\rm Tr}\vert K\vert^4\bigr)^{\perp}
\sum_{m,n} \Bigl\vert {\partial g\over \partial H_{mn}}\Bigr\vert^2
,\eqno(29)
$$
provided that non of the matrix algebras have dimension one.
In equation (29) we have denoted
$$\bigl({\rm Tr}\vert K\vert^4\bigr)^{\perp}={\rm Tr}(\vert K\vert^4)
-({\rm Tr}\vert K\vert^2 )^2 .\eqno(30)
$$
The fact that there are no contributions to the scalar potential
from the diagonal parts of the curvature is not very desirable, as
there will be many Higgs fields whose vevs are left undetermined.
To avoid this and to make sure that the potential has no flat directions
we consider the
following situation. By imposing a permutation symmetry
on two points (e.g. $ 1\leftrightarrow 2 $), we
can have  Higgs fields belonging to the adjoint representation
of the gauge group. This makes it possible to impose more constraints
on the mass matrices.
This case is very important for model building as
adjoint representations  are important
for symmetry breaking at high energies.
{}From the permutation symmetry we deduce that $M_{12}=M_{21}$, and
$a_1=a_2$. This in turn implies that $A_{\mu 1}=A_{\mu 2}$ and
$H_{12}=H_{21}=\Sigma $. It follows that
$$
H_{12}H_{21} \rightarrow g_1 H_{12}H_{21}g_1^{\ast} ,
$$
where we have used $g_1=g_2$.  With this it is now possible to
impose conditions of the form
$$
\sum_{p\not= m} M_{1p}M_{p1} =\alpha_{1} M_{12} +\beta_{1} 1_{n_1} .\eqno(31)
$$
Such  conditions could be generalised whenever two points
m and n have a permutation symmetry. In this case we write
$$
\sum_{p\not= m} M_{mp}M_{pm} =\alpha_{m} M_{mn} +\beta_m 1_{n_m}.
\eqno(32)
$$
Using equations (18) and (31) in (12) it is easily verified that $X_{mm}'$
simplifies to
$$\eqalign{
X_{mm}'+\sum_{p\not =m}\vert M_{mp}\vert^2 &=\sum_i a_m^i \di b_m^i
+(\partial^{\mu}A_{\mu m} +A^{\mu m}
A_{\mu m}) \cr
&\qquad  +\alpha_m \Sigma_m +\alpha_m \Sigma_m +\beta_m 1_{n_m},\cr} \eqno(33)
$$
where we have denoted $H_{mm}$ by $\Sigma_m $. This implies that
the diagonal components of $v_{mm}$, after projecting the
kernel of the Dirac operator, reduce to
$$
v_{mm}=\bigl( \vert K\vert^2\bigr)^{\perp} \Bigl(
\sum_{p\not= m}\vert H_{mp}\vert^2 -\alpha_m \Sigma_m -\beta_m 1_{n_m}
\Bigr) . \eqno(34)
$$
The contributions of the diagonal parts of the curvature to
the potential will be the square of such terms. The presence of
the new terms in the potential is a reflection of the fact that new terms to
the prepotential could be added. The most general gauge
invariant prepotential in this case is given by
$$\eqalign{
g&={\rm tr} \bigl( {1\over 3} (\sum_m\Sigma_m^3 +
\sum_{p\not= m, n}H_{mp}H_{pn}H_{nm})\cr
& -{1\over 2}\sum_m (\alpha_m \Sigma_m^2
+\sum_{p\not= n} \alpha_{mp}H_{mp}H_{pm}) -\beta_m \Sigma_m
\bigr) .\cr}
 \eqno(35)
$$
We are now in a position to write down the most general
physically acceptable potential that results from the
Yang-Mills noncommutative action. Assuming we have s of
the $n_m\times n_m $ matrices to be of dimension one, and
r couple of points have permutation symmetry, the scalar potential
will be given by
$$\eqalign{
V&= {\rm Tr}\bigl(\vert K\vert^4\bigr)^{\perp}
 \Bigl( \sum_s \bigl( \sum_p(\vert H_{sp}\vert^2 -\vert M_{sp}
\vert^2 )\bigr)^2 \cr
& \qquad \qquad + \sum_r \bigl( \vert \Sigma_r\vert^2 -\alpha_r \Sigma_r
-\beta_r \Sigma_r +\sum_{p\not= s} \vert H_{rp}\vert^2 \bigr)^2
\Bigr) \cr
 & +{\rm Tr}\vert K\vert^4 \sum_{m,n\not= r}\sum_{p\not= m, n}
\vert H_{mp}H_{pn}-\alpha_{mn}H_{mn}\vert^2 .\cr}\eqno(36)
$$
Naturally this formula could be reexpressed in terms of the
gauge invariant prepotential $g$ :
$$
V={\rm Tr}\bigl(\vert K\vert^4\bigr)^{\perp}\sum_r\Bigl\vert {\partial g
\over \partial \Sigma_r} \Bigr\vert^2 +{\rm Tr}\vert K\vert^4
\sum_{p\not= r}\Bigl\vert {\partial g\over \partial
H_{mp}}\Bigr\vert^2 , \eqno(37)
$$
The fact that the potential is a sum of squares
and is derivable from a prepotential makes it as near to
supersymmetric theories as possible without requiring supersymmetric
partners. However, they are not in general supersymmetric,
except in certain cases related to
$N=2$ and $N=4$ supersymmetry [9]. This, however, could be a signal
of the presence of new kind of symmetries, such as quantum
symmetries [8].

We conclude by noting that this derivation takes the pain of
obtaining the general form of the lagrangian for a unified
theory based on noncommutative geometry. Once the spectral
triple and the symmetries of the matrix algebras are specified,
the potential could immediately be written. This will make
constructing new realistic models based on noncommutative geometry
easier to achieve.
It is quite important to derive the new models using the anticipated
quantum
symmetries, in such a way that any predictions  made
could be protected at the quantum level by these  symmetries.
\vskip1truecm
{\bf\noindent Acknowledgments}\hfill\break
\vskip.2truecm
I would like to thank Alain Connes, J\"uerg Fr\"ohlich,
Olivier Grandjean, Daniel Kastler and Daniel Testard
for stimulating discussions. I would
also like to thank the Erwin Schr\"odinger International Institute
for Mathematical Physics, Vienna, for hospitality
where this work was done.
\vskip0.8truecm
{\bf \noindent References}
\vskip.2truecm
\item{[1]} A. Connes, {\sl Publ. Math. IHES} {\bf 62}  (1983) 44;\br
in {\sl the interface of mathematics
and particle physics }, Clarendon press, Oxford 1990, Eds
D. Quillen, G. Segal and  S. Tsou;\br
{\sl Noncommutative Geometry}, Academic Press, N.Y. (1994).

\item{[2]} M. Dubois-Violette, {\sl C.R. Acad. Sci. Paris},
{\bf 307} (1988) 403;\br
M. Dubois-Violette, R. Kerner and J. Madore, {\sl J. Math. Phys.}
{\bf 31} (1990) 316; {\sl Class. Quant. Grav.} {\bf 6} (1989) 1709.

\item{[3]}A. H. Chamseddine and J. Fr\"ohlich {\sl in Yang-
Festschrift} editors C. S. Liu and S.-T. Yau.

\item{[4]} A. Connes and J. Lott,{\sl Nucl.Phys.B Proc.Supp.}
{\bf 18B}  (1990) 29, North-Holland, Amsterdam;\br
{\sl in Proceedings of the 1991 Summer Carg\`ese Conference}
edited by J. Fr\"ohlich et al (Plenum, New York, 1992).

\item{[5]} D. Kastler, {\sl Rev. Math. Phys.} {\bf 5} (1993)
477;\br
D. Kastler and T. Sch\"ucker, {\sl Theor. Math. Phys.} {\bf
92} (1992) 522;\br
B. Iochum, D. Kastler and T. Sch\"ucker, Marseille preprint,
CPT-95/P.33197, and hep-th/9507150.

\item{[6]} A. H. Chamseddine, G. Felder and J. Fr\"ohlich,
{\sl Phys. Lett.} {\bf B296} (1992) 301; {\sl Nucl.Phys.}
{\bf B395} (1993), 672;\br
A. H. Chamseddine and J. Fr\"ohlich, {\sl Phys. Rev.} {\bf D50}
(1994) 2893.

\item{[7]}E. Alvarez, J. M. Garcia-Bondia, and C. P. Martin,
{\sl Phys. Lett.} {\bf B306} (1993) 55.

\item{[8]}A. Connes, {\sl Plenary Talk Presented at the Trest Meeting
on Noncommutative Geometry} May 1995.

\item{[9]}A. H. Chamseddine {\sl Phys. Lett.} {\bf B332} (1994) 349.
\end